\documentclass[sigconf,nonacm]{acmart}

\makeatletter
\def\ACM@cc@type{by} 
\makeatother

\AtBeginDocument{%
  }

\begin{document}

\title{A Safety-First Gateway Architecture for Trusted Public Health Resource Navigation}

\author{Ben Torkian}
\authornote{Corresponding author.}
\email{torkian@email.sc.edu}
\orcid{1234-5678-9012}
\affiliation{%
  \institution{University of South Carolina}
  \city{Columbia}
  \state{South Carolina}
  \country{USA}}

\author{Jun Zhou}
\affiliation{%
  \institution{University of South Carolina}
  \city{Columbia}
  \country{USA}}
\email{zhouj@mailbox.sc.edu}

\renewcommand{\shortauthors}{Torkian et al.}

\begin{abstract}

Conversational AI can improve access to public health information, but
public-facing healthcare applications require safeguards against inappropriate
medical guidance and unsupported generation. We present a Safety-First Science
Gateway for maternal and child health (MCH) resource navigation that combines
large language models (LLMs) and retrieval-augmented generation (RAG) with a
multi-layer safety architecture. The gateway integrates emergency handling,
domain/scope screening, source attribution, anonymous session management, and
operational audit logging while restricting retrieval to curated institutional
resources. We describe the gateway architecture, prototype implementation, and
functional verification of selected workflows. The current system provides
resource provenance and safety-bounded navigation; it does not constitute a
clinical decision-support system or automated claim-by-claim verification of
generated health information. This work provides a reusable architectural
framework for conversational navigation of curated public-health resources.

\end{abstract}

\ccsdesc[500]{Applied computing~Health care information systems}
\ccsdesc[500]{Computing methodologies~Information extraction}
\ccsdesc[500]{Human-centered computing~Natural language interfaces}
\ccsdesc[500]{Computer systems organization~Cloud computing}


\maketitle

\section{Introduction}
Public health organizations maintain extensive collections of online resources, but these resources are often distributed across multiple agencies and websites, making them difficult to discover and navigate. Conversational AI offers a natural interface for accessing such information, while retrieval-augmented generation (RAG) enables responses to be grounded in curated knowledge sources rather than relying solely on large language models (LLMs) \cite{lewis2020retrieval}. However, healthcare applications require additional safeguards because LLMs may generate unsupported medical information or recommendations beyond their intended scope \cite{meyrowitsch2023misinformation, sittig2025recommendations}.

Existing healthcare conversational systems have been used for patient education, behavioral support, and care management, but their deployment continues to raise safety, privacy, and evaluation challenges \cite{laymouna2024chatbots,kocaballi2022design}. In maternal and child health, Nguyen et al. developed and evaluated \emph{Rosie}, a question-and-answer chatbot that provides new mothers with information from a curated knowledge base of vetted sources \cite{nguyen2024rosie}. Building on this landscape, our work focuses on a different problem: governed navigation of a locally curated resource collection. Rather than providing diagnosis, triage, treatment, or individualized clinical advice, the proposed gateway combines source-constrained retrieval with provenance, safety boundaries, and operational monitoring to support public-facing access to South Carolina MCH resources.

This paper presents a Safety-First Science Gateway for maternal and child health (MCH) resource navigation. The gateway combines a controlled RAG framework with a multi-layer safety architecture that includes domain validation, emergency detection, citation enforcement, anonymous session management, and operational audit logging. 
Rather than relying primarily on the LLM's pretrained medical knowledge, responses are constrained to institutionally curated MCH resources, providing transparent and traceable access to trusted public health information. Unlike standalone conversational agents, the proposed gateway integrates AI services, curated public health resources, operational governance, and monitoring into a unified platform for trusted public access.

The main contributions of this work are:
\begin{itemize}
    \item A safety-first gateway architecture that integrates conversational AI with layered safeguards for trustworthy public health resource navigation.
    \item A controlled RAG framework that constrains responses to curated institutional resources while preserving information provenance.
    \item A deployed prototype implementation supporting anonymous multi-user sessions, operational monitoring, and audit logging.
\end{itemize}

Although developed for maternal and child health, the proposed architecture is applicable to other domains requiring trustworthy conversational access to curated institutional knowledge. 

\section{Gateway Architecture}
\label{sec:architecture}

\begin{figure*}[htp]
\centering
\includegraphics[width=\linewidth]{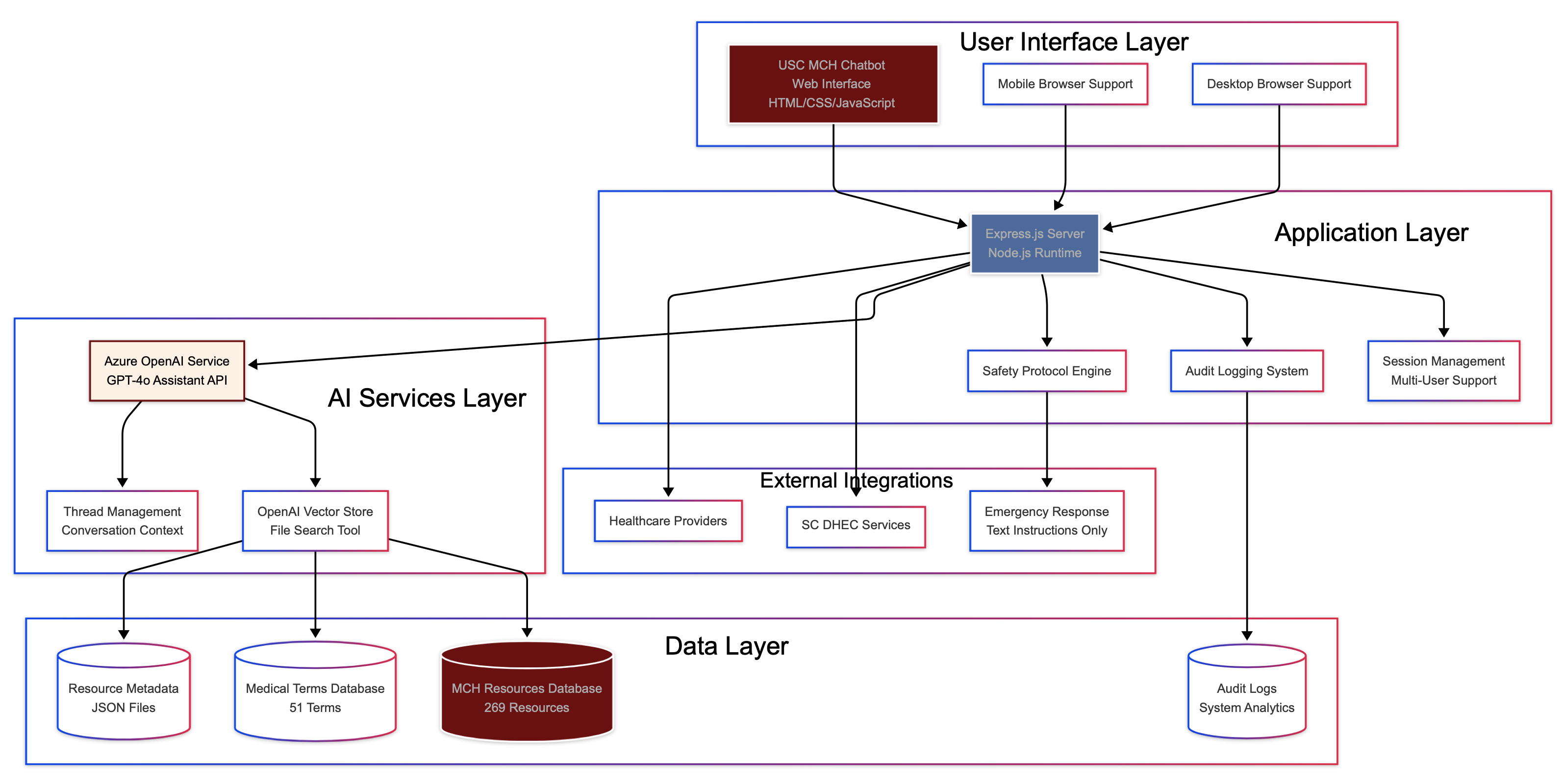}
\caption{High-level system architecture.}
\label{fig:IFS_architecture}
\end{figure*}

The proposed gateway adopts a layered architecture that separates user interaction, application services, AI services, and data management, as shown in Fig.~\ref{fig:IFS_architecture}. The user interface provides a web-based conversational experience, while the application layer orchestrates request processing, safety enforcement, and response generation. The AI services layer supports RAG over the curated knowledge repository. The data layer manages both the curated knowledge repository used for retrieval and the operational data supporting monitoring and analytics. This separation decouples gateway policies from the underlying language model, allowing operational functions such as safety validation, session management, and audit logging to evolve independently of the AI service.

A key design objective is to provide traceable access to curated public-health
resources while limiting inappropriate use of generative AI. Rather than relying
primarily on the LLM's internal knowledge, the gateway retrieves curated MCH
resource records through a controlled RAG pipeline \cite{lewis2020retrieval}.
The records include metadata for retrieval, categorization, and source
attribution. Responses must identify the retrieved resources used for
navigation, and the system returns a safe fallback when retrieval does not
provide relevant material. These mechanisms provide provenance and reduce
unsupported generation; however, the prototype does not yet perform automated
claim-by-claim verification that generated statements are entailed by retrieved
content. Source attribution should therefore be understood as traceability, not
as a guarantee of factual or clinical correctness.

\begin{figure*}[htp]
  \centering
  \includegraphics[width=0.9\linewidth]{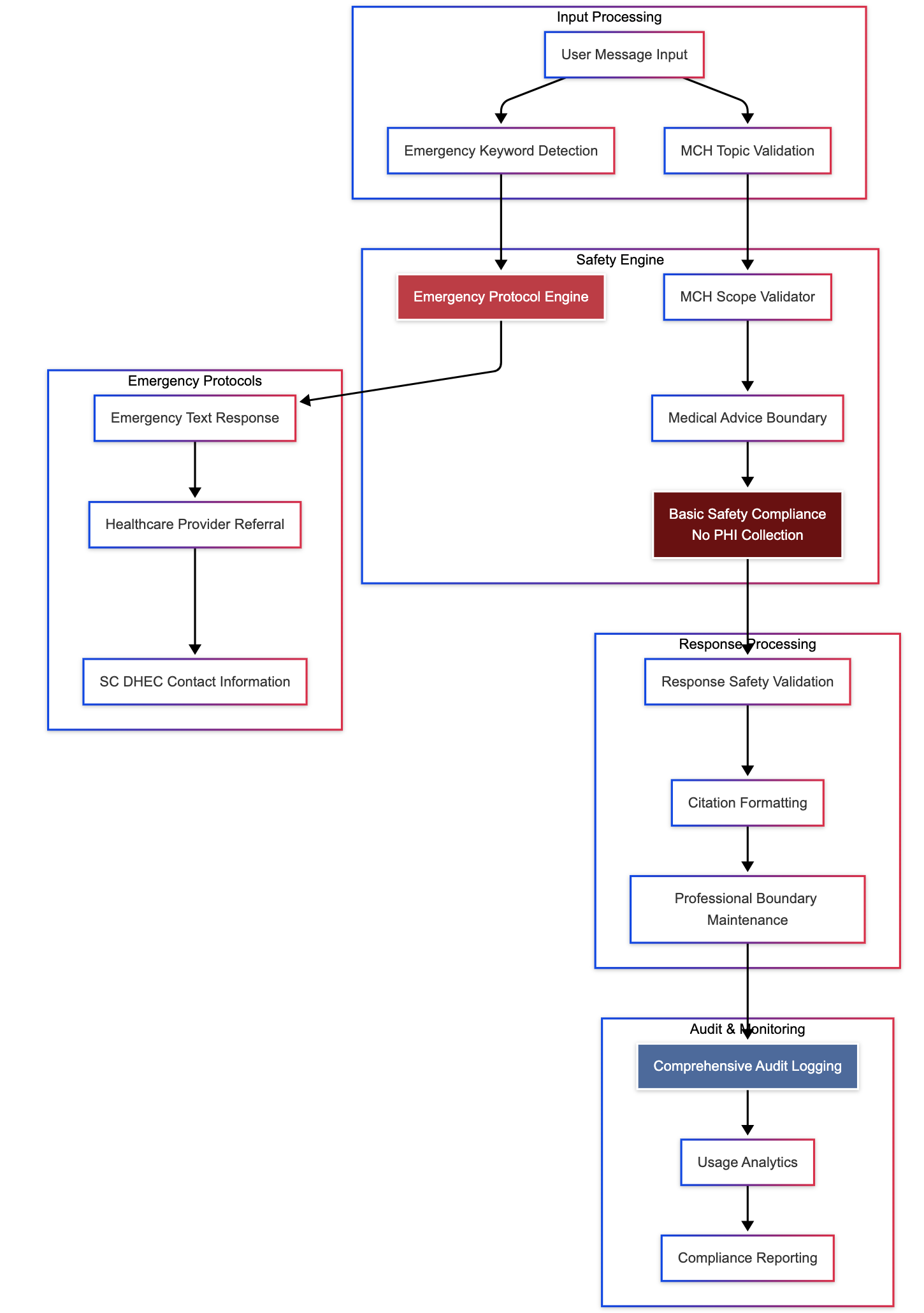}
  \caption{Safety and Compliance Architecture detailing the multi-layered safety engine, emergency detection, and audit logging systems.}
  \label{fig:IFS_audit}
  \vspace{-1em}
\end{figure*}

To support safe public-facing deployment, the gateway incorporates a multi-layer safety architecture, illustrated in
Fig.~\ref{fig:IFS_audit}. The safety policy includes emergency-language handling and domain/scope validation alongside the standard retrieval workflow. The response stage is limited to curated resources and requires source attribution; defined checks then apply
the medical-advice and professional-role boundaries before delivery. The
medical-advice boundary excludes diagnosis, treatment recommendations,
medication or dosage guidance, and individualized clinical decisions.
Professional-boundary maintenance requires the gateway to identify itself as a
resource-navigation tool and direct users to qualified professionals or
emergency services when appropriate.

Emergency-language handling is currently implemented through response-generation instructions and a predefined response policy. Anonymous session identifiers preserve conversational context without account creation, but do not prevent users from entering personal health information in free-text prompts. The interface therefore instructs users not to enter identifying information. The system distinguishes public resource records, Azure-hosted retrieval artifacts, ephemeral in-memory session mappings, and operational logs; it does not store patient or clinical records. Azure-hosted retrieval artifacts and conversation threads are protected through the platform's encryption at rest and key-based access controls, whereas local audit logs remain the primary unaddressed protection boundary in the current prototype. Structured audit logs capture system events, safety decisions, and operational metrics to support monitoring and debugging. Although anonymous sessions avoid account creation, users may still enter personal health information in free-text prompts. The current prototype does not implement prompt redaction, retention
limits, or formal log-access policies; these are important requirements for a future production deployment.

\section{Implementation}
\label{sec:implementation}
The proposed architecture has been implemented as a deployed prototype web application. The gateway uses a Node.js/Express backend and a lightweight HTML/CSS/JavaScript frontend. Conversational responses are generated using Azure OpenAI GPT-4o with RAG, while Azure AI Search provides semantic retrieval over a curated repository of MCH resources. The repository includes institutionally curated MCH resources and associated medical terminology, enriched with metadata to support retrieval and source attribution. The implementation also incorporates anonymous session management, structured JSON audit logging, and environment-based configuration to support deployment, monitoring, and maintenance.

The gateway is a public, read-only service for navigating curated maternal and
child health (MCH) resources relevant to South Carolina residents. Its scope is
limited to locating and presenting resources contained in the curated repository;
it does not provide diagnosis, triage, treatment, or individualized medical
advice. The corpus contains 269 public resource records and 51 medical terms, with
structured metadata including source organization, title, URL, theme, audience,
format, language availability, and reviewer comments. Public users access the
interface anonymously. Resource curation and system management occur out of
band through institutional Azure credentials and environment-based
configuration; the current prototype does not implement in-application roles or
RBAC for data providers, curators, or administrators.

Trust is addressed at both source and resource levels, but is not formally
certified by the prototype. The resource list was compiled by the Institute for
Families in Society and underwent item-level review and metadata annotation with
domain-expert input. Each indexed record preserves source attribution for user
inspection. Functional verification evaluates retrieval and attribution
behavior; it does not independently verify organizational authority or the
accuracy, completeness, or currency of every resource. Formal inclusion
criteria, periodic revalidation, and resource-withdrawal procedures remain
future governance work.

\section{System Verification}
\label{sec:validation}
The gateway was functionally verified using representative operational
scenarios. Each scenario was assessed against an observable acceptance criterion
and recorded outcome (Table~\ref{tab:verification}). This verification evaluates
selected workflow behavior, not clinical safety, factual accuracy,
retrieval-quality, or real-world user outcomes.
\begin{table*}[htp]
\caption{Representative Functional Verification Scenarios}
\label{tab:verification}
\centering
\small
\begin{tabular}{p{2.8cm}p{5.8cm}p{1.0cm}}
\toprule
\textbf{Scenario} & \textbf{Acceptance criterion} & \textbf{Outcome} \\
\midrule
Resource request & Returns relevant curated resource link(s) with source attribution. & Met \\
Out-of-domain query & Does not provide an unsupported answer; returns an appropriate redirect. & Met \\
Emergency request & Returns the predefined emergency message according to the implemented emergency workflow. & Met \\
Multi-turn conversation & Retains context within a session and isolates separate sessions. & Met \\
Audit logging & Records defined system events, safety decisions, and operational metrics. & Met \\
\bottomrule
\end{tabular}
\end{table*}

\section{Design Considerations and Future Work}
\label{sec:discussion}

Designing conversational science gateways for public-facing healthcare applications requires balancing usability, safety, and operational simplicity. Rather than maximizing the generative capabilities of the underlying language model, our design emphasizes trustworthy access to curated public health resources through layered safety mechanisms, information provenance, and operational governance. Three key design decisions shaped the gateway architecture. First, grounding responses in curated resources improves traceability and reduces reliance on the LLM's internal knowledge, although it requires ongoing maintenance of the knowledge repository. Second, separating safety enforcement from response generation enables gateway policies to be applied consistently, independent of the underlying AI model. Finally, anonymous session management and structured audit logging support operational monitoring. However, anonymous sessions do not prevent users from entering personal health information in free-text prompts; the current prototype does not yet implement prompt redaction, retention limits, or formal log-access
policies.

Although this work focuses on maternal and child health resource navigation, the proposed architecture is applicable to other domains requiring conversational access to trusted public health resources under strict safety and governance requirements.

The current work focuses on gateway architecture and implementation rather than large-scale evaluation. Future work will include longitudinal deployment to collect operational usage data, quantitative evaluation of retrieval effectiveness and user experience, expansion of multilingual support, and continuous refinement of the curated knowledge repository. These efforts will provide additional evidence of the gateway's effectiveness while further improving accessibility and scalability.
Future work will also investigate automated evidence--claim alignment, formal
content-governance procedures, broader controls for prompt-injected PHI, and
evaluation of retrieval quality and user experience.

\section{Conclusion}
\label{sec:conclusion}
We presented a safety-first gateway architecture that combines conversational AI, controlled retrieval, layered safety enforcement, and operational governance for public health resource navigation.
By combining curated retrieval, visible source provenance, safety-bounded
response policies, and operational monitoring, the architecture supports
traceable public-facing navigation of institutional resources while maintaining transparency and information provenance. Although demonstrated in maternal and child health, the architectural principles are broadly applicable to other domains that require safe and reliable conversational access to institutional knowledge.

\textbf{System Availability.}
A live deployment is available at \url{https://empower-assistant.sc.edu/}. The source code is not publicly released because the system includes institution-specific configurations and curated resources; additional implementation details are available from the corresponding author upon reasonable request.

\bibliographystyle{ACM-Reference-Format}
\bibliography{IFS}

\end{document}